\documentclass[10pt, groupedaddress, twocolumn]{revtex4-1}
\usepackage[T1]{fontenc}
\usepackage{bm}
\usepackage{graphicx}
\usepackage{amsmath}
\usepackage{amssymb}
\usepackage{sidecap}
\usepackage{dcolumn}
\usepackage{color}
\usepackage{epstopdf}
\usepackage{ifpdf}
\usepackage{sidecap}
\newcommand{\eps}{\varepsilon}

\begin{document}
\title{\textbf{Characterization of the size and position of electron-hole puddles at a graphene p-n junction}}
\date{Antwerp, \today}

\author{S. P. Milovanovi\'{c}} \email{slavisa.milovanovic@uantwerpen.be}
\affiliation{Departement Fysica, Universiteit Antwerpen, \\
Groenenborgerlaan 171, B-2020 Antwerpen, Belgium}

\author{F. M. Peeters}\email{francois.peeters@uantwerpen.be}
\affiliation{Departement Fysica, Universiteit Antwerpen, \\
Groenenborgerlaan 171, B-2020 Antwerpen, Belgium}
\date{\today}

\begin{abstract}
The effect of an electron-hole puddle on the electrical transport when governed by snake states in a bipolar graphene structure is investigated. Using numerical simulations we show that information on the size and position of the electron-hole puddle can be obtained using the dependence of the conductance on magnetic field and electron density of the gated region. The presence of the scatterer disrupts snake state transport which alters the conduction pattern. We obtain a simple analytical formula that connects the position of the electron-hole puddle with features observed in the conductance. Size of the electron-hole puddle is estimated from the magnetic field and gate potential that maximizes the effect of the puddle on the electrical transport.
\end{abstract}

\maketitle
\section{Introduction}
Graphene is expected to have a large impact on future electronics due to its reduced dimensionality, linear band structure, high electron and thermal conductivity \citep{r1}. Furthermore, by applying gates below and above the graphene sample it is possible to form n-n, p-n, and p-p junctions within a single sheet of material \citep{r10}. Owing to the zero energy gap and the relativistic nature of the carriers graphene p-n junctions exhibit new and exciting physical features, unobserved in semiconducting electronics, e.g. Klein tunnelling \citep{r11, r12}, valley-valve effect \citep{r13}, and snake states \citep{r90, r100, r101, r102}, to name a few. Recently, encapsulating graphene in h-BN \citep{r49, r50, r51, r52} has become a popular technique for the realization of high-quality graphene devices. Encapsulation enables one to make flat, ultra clean graphene surfaces resulting in a mean free path for the carriers exceeding the dimensions of the system \citep{r50}. Furthermore, the top h-BN layer can serve as a dielectric for local gating of the graphene underneath. This allows the realization of very sharp potential profiles which is a prerequisite for the observation of various phenomena in graphene, including Klein tunnelling \citep{r60}, Veselago lensing \citep{r70, r71}, quantum interference \citep{r80}, etc. 

It was shown that the presence of the local gate leads to the appearance of electron-hole (e-h) puddles along the p-n interface \citep{r90, r91, r92, r93}. Up to now, the only way to observe these scatterers was by using different scanning techniques \citep{r91, r92}. Unfortunately, this is often not practical or in the case of suspended devices almost impossible. However, at a p-n interface such e-h puddles have a large impact on electrical transport \citep{r102, r90, r101} that occurs predominantly through snake states. In order to quantify such e-h puddles we investigate the influence of these scatterers on the transport due to snake states  along the p-n interface in the bipolar regime. In this study we show that from the magnetic field and density dependence of the conductance we can extract information on the size and position of the e-h puddles that are near the p-n interface.
\begin{figure}[htbp]
\begin{center}
\includegraphics[width=6.0cm]{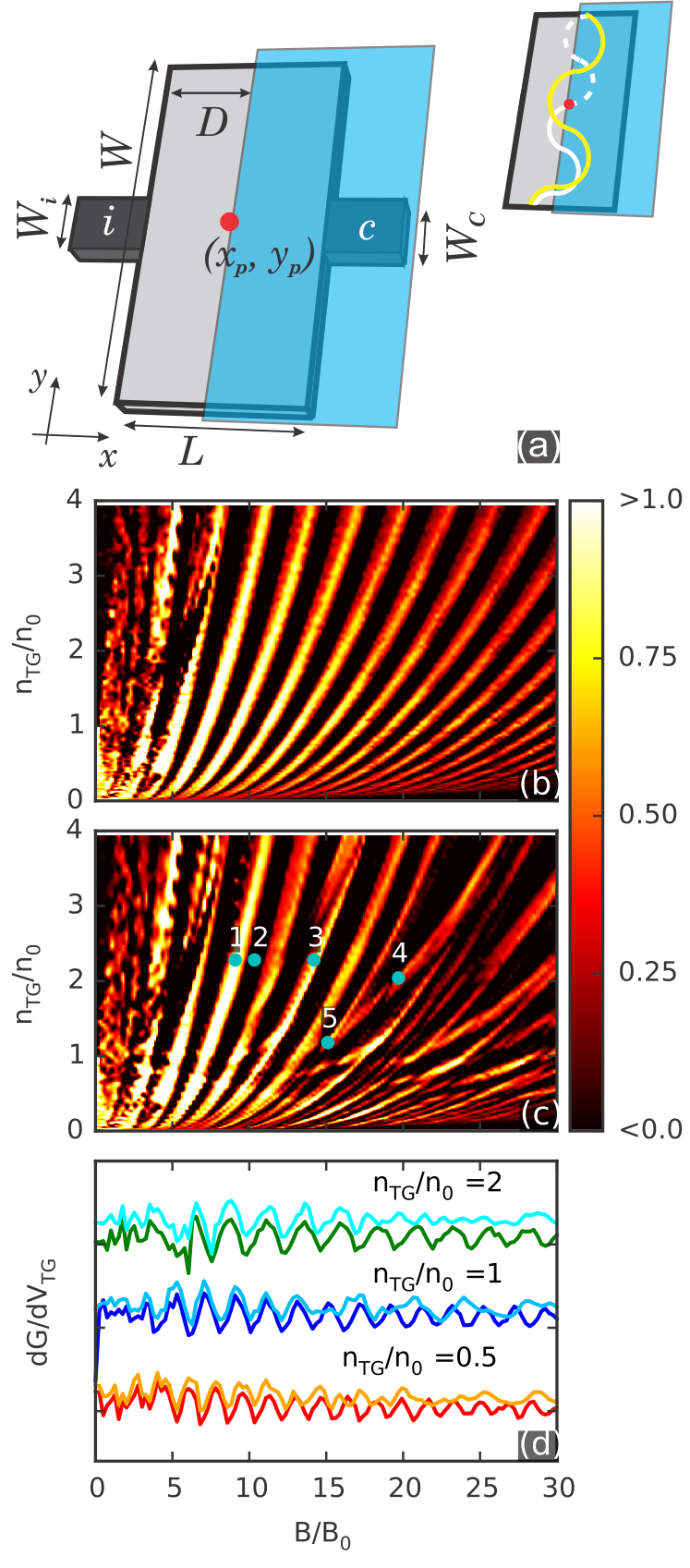}
\caption{(Color online) (a) Schematics of the device. The e-h puddle (red circle) is positioned at $(x_p, y_p)$. The right inset shows a sketch of snake states colliding with  the e-h puddle (white curve) and flowing around it (yellow curve). (b) Contour plot of $dG/dV_{TG}$ versus applied magnetic field and density of the gated region in the case when there are no e-h puddles in the system and (c) in the presence of one e-h puddle positioned at ($x_p, y_p, r_p$)=(0.48, 0.75, 0.05). (d) Cuts of $dG/dV_{TG}$ at constant $n_{TG}$ are shown for the system without (darker color scheme) and with an e-h puddle (bright color scheme). The curves are shifted for clarity.} \label{f0}
\end{center}
\end{figure}
\section{Methods}

We simulate a graphene system of size $L \times W$ with a top gate that controls the density of the graphene sample, as shown in Fig. \ref{f0}(a). The p-n interface is positioned at $D = L/2$. Electrons are injected from a narrow injector placed in the non-gated part of the device and collected at the collector placed under the top gate. Because snake states are classical objects it is sufficient to simulate transport using the well-known semiclassical billiard model \citep{r102, r110} in combination with the Landauer formula. To simplify calculations the following dimensionless units are used: $l \rightarrow l/l_0$ where $l = \{L, W, D, W_{i, c}\}$, and $l_0$ is the unit of length, $\eps \rightarrow \eps/\eps_0$ where $\eps = \{E_F, eV_{TG}\}$ and $\eps_0$ is the unit of energy, and $B \rightarrow B/B_0$ where $B_0 = \eps_0/(ev_F l_0)$. For $\eps_0 = 100$ meV and $l_0 = 1$ $\mu$m this will result in $B_0 = 0.1$ T. In the  following we used:  $E_F = 1$, $L = 1$, $W = 2$, D= 0.5, and $W_{i,c}$ = 0.3 is the width of the injector/collector which corresponds to typical experimental systems. Furthermore, instead of using the dependence of the conductance on the value of the applied potential we will use the electron density of the gated region, $n_{TG}$, given by
\begin{equation}\label{entg}
n_{TG} = s \frac{(E_F-eV_{TG})^2}{\pi\hbar^2v_F^2},
\end{equation} 
with $s = sign(E_F-eV_{TG})$, or in dimensionless units $n_{TG} \rightarrow n_{TG}/n_0$ where $n_0 = \eps_0^2/(\pi\hbar^2v_F^2)$. The two-terminal conductance is given in units of $4e^2/h$, where the factor 4 is due to spin and valley degeneracy.

It is expected that e-h puddles are localized around the p-n interface and appear due to the local change of the gating potential rather than they are a consequence of localized charges induced by the substrate. Following Ref. \onlinecite{r90} the interaction of the current carriers with the puddle is modelled as follows. Whenever an electron appears within a radius $r_p$ of the center of the puddle positioned at ($x_p, y_p$), it is moved to the center of the puddle and the direction of the velocity vector is taken randomly. Such an approach takes into account the irregular edges of the e-h puddle.
\section{Results and discussion}

\begin{figure*}[htbp]
\begin{center}
\includegraphics[width=18cm]{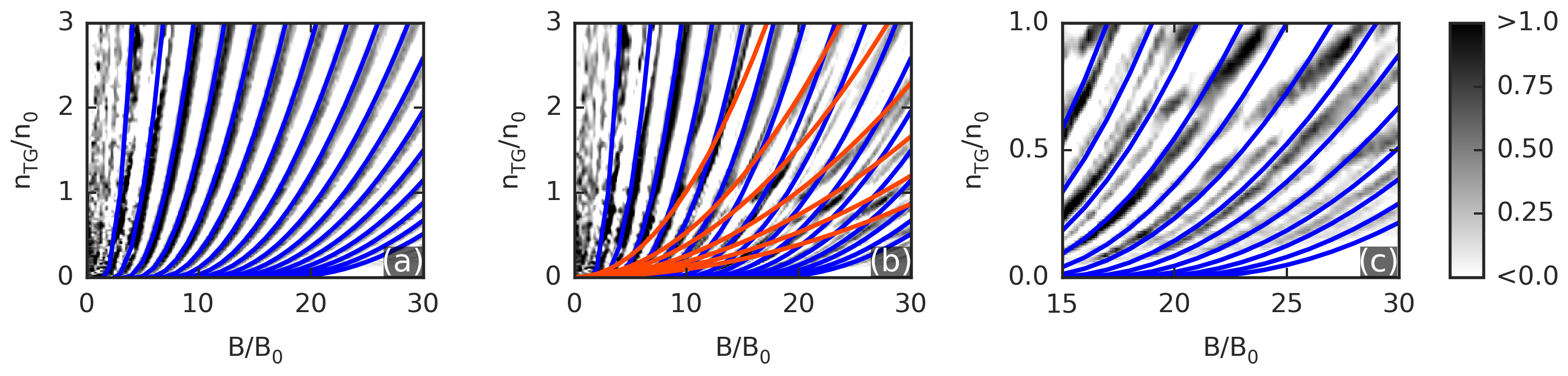}
\caption{(Color online) (a) The same as Fig. \ref{f0}(a) but where now the blue curves show the peak positions as given by Eq. \eqref{e1}  with $L_{PN} = 2$. (b) The same as Fig. \ref{f0}(b) where in addition the results from Eq. \eqref{e3} are supperimposed as red curves with $r_{TG}$: 0.1, 0.072, 0.061, 0.05, 0.042, 0.036, and 0.03. (c) The same as Fig. \ref{f0}(b). Blue curves are the analytical results from Eq. \eqref{e1}  but now with $L_{PN} = 1.25$. } \label{f1}
\end{center}
\end{figure*}

Figs. \ref{f0}(b) and (c) show the derivative of the conductance with respect to the gate potential versus the magnetic field and density of the gated region for the case without and with an e-h puddle, respectively. In the latter case the e-h puddle is placed near the p-n interface with the following parameters ($x_p, y_p, r_p$) = (0.48, 0.75, 0.05), where $x_p$ and $y_p$ are the scaled $x$- and $y$-coordinate of the center and $r_p$ is the radius of the puddle. The size of the puddle is of the order observed in experiments \citep{r51, r91}. If there is no puddle in the system, the derivative of the conductance exhibits regular oscillations, as shown in Fig. \ref{f0}(b). This is also seen in Fig. \ref{f0}(d) where we show cuts of $dG/dV_{TG}$ for constant $n_{TG}$. 
The maxima in the conductance are determined by
 \begin{equation}\label{e1}
 n_{TG}^n = \left(\frac{L_{PN}}{2n+ 1}B\right)^2 ,
\end{equation}  
where $L_{PN}$ is the length of the p-n interface and $n$ is an integer. This equation is derived from
 \begin{equation}\label{e2}
 L_{PN} = (2n + 1)(r_{TG} + r_{BG}),
\end{equation}  
where $r_{TG}$ and $r_{BG}$ are the cyclotron radius in the gated and non-gated region, respectively. Eq. \eqref{e1} is based on transport due to snake states along the p-n interface which, together with Klein tunnelling in graphene, results in oscillations of the conductance \citep{r90, r101, r102}. Namely, depending on the value of the magnetic field, the Fermi energy, and applied potential the injected electron beam can be directed towards injector or collector. If the length of the p-n interface is equal to an odd (even) number multiple of $r_{TG} + r_{BG}$ the injected beam will be directed towards the collector (injector) and the conductance will exhibit a maximum (minimum). In Fig. \ref{f1}(a) we show that the numerical results are in agreement with the analytical predictions (blue curves) confirming that transport is governed by snake states.

Fig. \ref{f0}(c) shows the derivative of the conductance when an e-h puddle is introduced near the p-n interface. The periodicity of the pattern is partially destroyed as seen in this figure as well as in Fig. \ref{f0}(d) where we show cuts of the conductance for certain values of $n_{TG}$. The reason behind this is as follows. Depending on the value of the cyclotron radius in the gated and non-gated region injected electron beam can flow around the e-h puddle without being (or very little) affected by its presence, as shown in the inset of Fig. \ref{f0}(a) by the yellow curve. Hence, the conductance will not change. If, on the other hand, the injected beam collides with the puddle, as shown in the inset of Fig. \ref{f0}(a) by the white curve, its presence will alter the snake state trajectory and the oscillatory pattern of the conductance will change. However, the following new features are observed in Fig. \ref{f0}(c): 1) crossings of curves, and 2) new maxima that do not follow Eq. \eqref{e1}. Similar features were observed in the experiment of Ref. \onlinecite{r90}. Since we only added an e-h puddle to the system these features are a consequence of the interaction between the snake states and the e-h puddle and therefore they can be used to gain information about the size and position of the puddle.

The cyclotron radius in the gated region depends on the magnetic field and the carrier density. In dimensionless units this is given by,
 \begin{equation}\label{e3}
 r_{TG} = \frac{\sqrt{n_{TG}}}{B}.
\end{equation}  
Hence, by increasing (decreasing) the density (magnetic field) we increase the cyclotron radius. Fig. \ref{f1}(b) shows the contour plot of $dG/dV_{TG}$ together with the results of Eq. \eqref{e1} plotted on top of it (blue curves). Note that Eq. \eqref{e1} agrees with the simulation in the upper part of the figure where $r_{TG}$ is large. However, as we decrease the radius (bottom part) other features become dominant. The reason for this is that the size of the cyclotron radius in both regions of the p-n junction become smaller or at least comparable with the radius of the puddle. This will have a significant effect on the trajectories of the snake states and alters the output signal. 
\subsection{Determining the position of the e-h puddle}

Interesting is the bottom right part of Fig. \ref{f0}(c). This is the region of small gate density and large magnetic field which means that the cyclotron radius in both regions of the p-n junction are small and the results from Eq. \eqref{e1} deviate most strongly from our numerical results. Furthermore, Eq. \eqref{e1} with $L_{PN} = 2$ suggests that this region should have many more peaks than observed. $r_{BG}$ and $r_{TG}$ in this  region are smaller than the radius of the puddle which implies that all electrons that reach the p-n interface will be scattered by the puddle. Thus, the puddle effectively shortens the p-n interface to $L_{PN}^* =  W - y_p$, which is the length of the p-n interface from the puddle. In Fig. \ref{f1}(c) we show the results of Eq. \eqref{e1} with $L_{PN} = W - y_p = 1.25$ by the blue curves which agree with the numerical data. In this way we are able to extract information about the position of the e-h puddle.
\begin{figure}[htbp]
\begin{center}
\includegraphics[width=8.5cm]{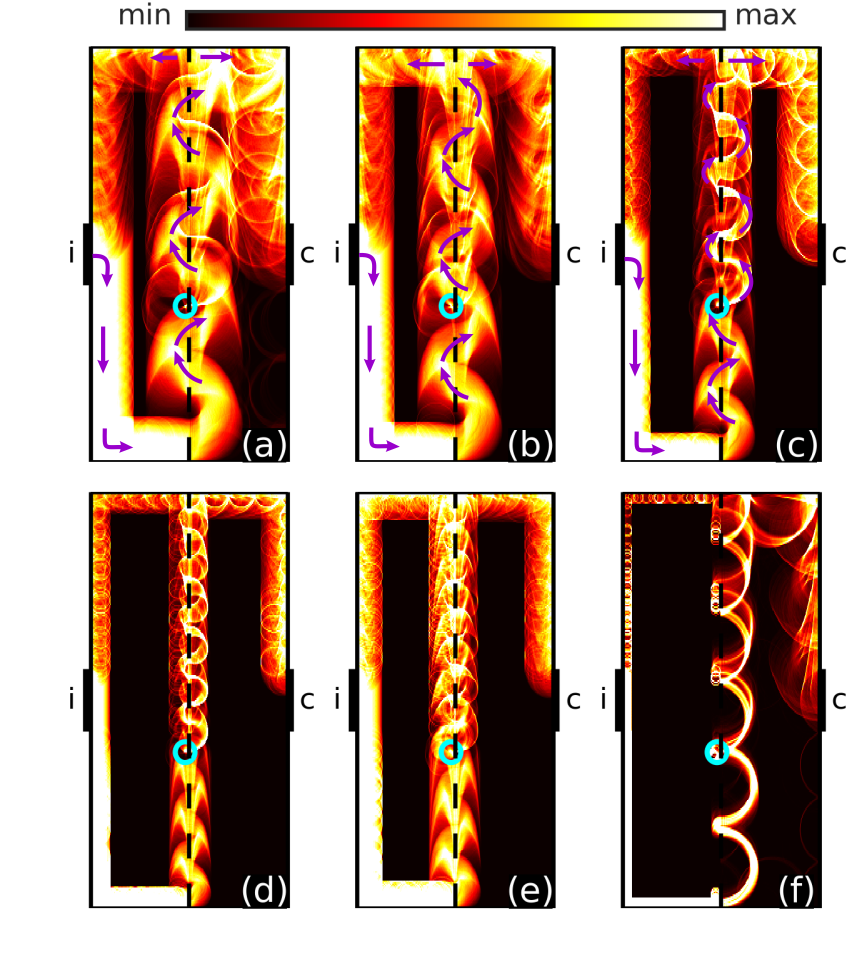}
\caption{(Color online) (a)-(e) Current density plots for the points 1-5 in Fig. \ref{f0}(b). The e-h puddle is shown by the cyan circle at the p-n interface while the arrows show the direction of the current. (f) Current density plot for $E_F = 0.25$, $n_{TG} = 2.5$, and $B = 12$.} \label{f2}
\end{center}
\end{figure}

Fig. \ref{f0}(c) shows interesting features in the high magnetic field region - curves that connect neighbouring curves that could be described by Eq. \eqref{e1}. Their origin can be traced back to the presence of disorder in the system. We find that these features satisfy Eq. \eqref{e3} for a specific value of $r_{TG}$. In other words, these curves connect points with the same cyclotron radius in the gated region. 
Using Eq. \eqref{e3} we took $r_{TG}$ as a fitting parameter which results in the red curves in Fig. \ref{f1}(b) for the following values of $r_{TG}$: 0.1, 0.072, 0.061, 0.05, 0.042, 0.036, and 0.03.

In order to better understand the latter features we use current density plots. The plots in Figs. \ref{f2}(a)-(e) show the flow of the current carriers in the system for the points 1-5 indicated in Fig. \ref{f0}(c). Notice the difference between Figs. \ref{f2}(a) and \ref{f2}(b). In both figures the injected beam flows around the puddle and only a small amount of carriers is affected by it. However, in Fig. \ref{f2}(a) we show the case when the transmission has a local maximum and in Fig. \ref{f2}(b) when the reflection has a local maximum. This is clearly seen if we follow the flow of the injected beam. In Fig. \ref{f2}(a) the beam injected at terminal $i$ bends to the right and hits the p-n interface from the bottom. We can follow its transmissions through the p-n interface until it finally ends up at the gated side of the p-n junction and is collected by the collector $c$. In Fig. \ref{f2}(b) we have a similar situation however now a larger part of the beam ends up at the non-gated side and is transported back to the injector. Fig. \ref{f2}(c) shows a similar situation as Fig. \ref{f2}(a) with the difference that the electron beam is now strongly affected by the e-h puddle. Notice the shift of the beam after it encounters the puddle. Similar shift can be observed in Fig. \ref{f2}(d) which is for point 4 in Fig. \ref{f0}(c) on the curve with constant $r_{TG} = 0.061$ (see also Fig. \ref{f1}(b)). This curve appears in the region where the first derivative is negative similar as for point 2 in Fig. \ref{f0}(b). In the latter case the electron beam is not affected by the puddle and it ends up at the injector's side. However, in Fig. \ref{f2}(d) we see that the part of the beam that is scattered by the puddle ends up at the collector side of the p-n junction. Therefore, this feature appears when the shift of the beam that interacts with the puddle is such that it changes the region in which the beam will end up. If we follow this curve of constant $r_{TG}$ eventually we will reach one of the snake state curves. This is shown in Fig. \ref{f2}(e) where we see that for this point a part of the injected beam is still affected by the puddle but the part that flows without interacting with it is much larger than in previous case.
\subsection{Determining the size of the e-h puddle}

Next, we will look for a way to obtain information about the size of the e-h puddle. A rough estimate can be made using previous results. Firstly, one needs to locate a region of $dG/dV_{TG}$ with the new features. Secondly, one draws a curve of constant $r_{TG}$ such that the located region lays below it. Since we know that the conductance is most strongly affected by the puddle if $r_{TG} < r_p$, chosen value of $r_{TG}$ can serve as a rough estimate of $r_p$.
Inspecting Figs. \ref{f1}(b) and (c) we find that this region lays below the $r_{TG} = 0.72$ curve and this can be a rough estimate for $r_p$. However, for a more precise estimate we proceed as follows. We decrease the back-gate voltage to a very low value, fix the magnetic field and change the top-gate voltage. The idea is to have a very small cyclotron radius in the non-gated region and a large $r_{TG}$. In this way the signal will be affected by the e-h puddle only when the part of the beam on the n-side goes around the puddle. This is shown in Fig. \ref{f2}(f). However, we should be careful that $2r_{TG}$ is never larger than the width of the collector otherwise carriers will not be collected and will propagate to the p-n interface. This can be controlled by the magnetic field. 
\begin{figure}[htbp]
\begin{center}
\includegraphics[width=8.5cm]{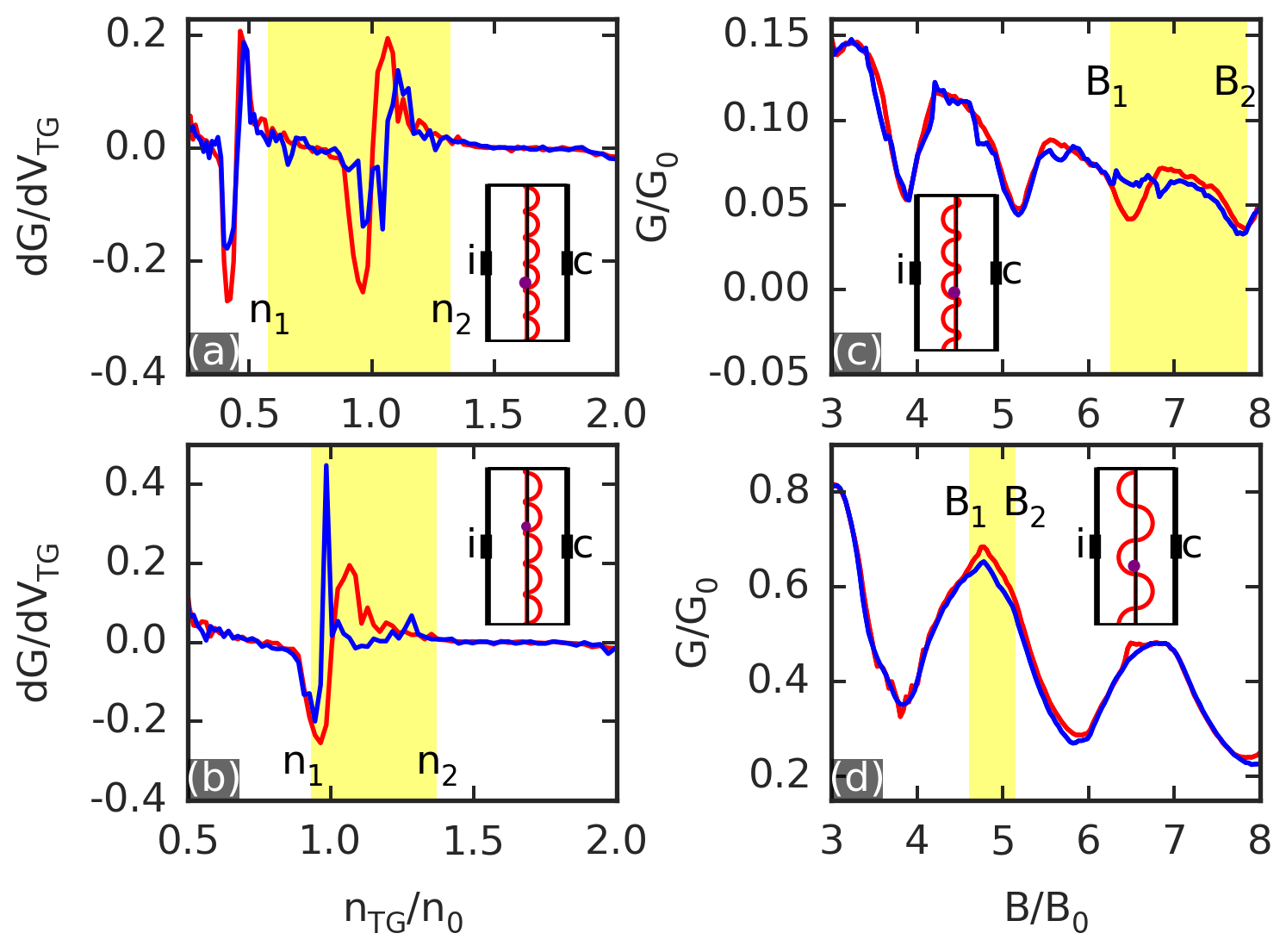}
\caption{(Color online) Derivative of the conductance for the case when there is no e-h puddle (red curve) and when there is one e-h puddle (blue curve) with parameters (a) ($x_p, y_p, r_p$) = (0.48, 0.75, 0.05) and (b) ($x_p, y_p, r_p$) = (0.49, 1.25, 0.035). Plots are made for $B = 10$ and $n_{BG}/n_{0} = 0.06$. (c) Conductance versus the magnetic field for the same puddle as in (a). The plot is made for $n_{TG} = 0.1$ and $n_{BG}/n_{0} = 0.06$. (d) The same as (c) but with $n_{TG} = 1$. The inset in each figure shows schematically the snake trajectory for point $n_1$ $(B_1)$ and the blue circle indicates the position of the puddle. The difference between the two curves in the yellow region is a consequence of the presence of the scatterer.} \label{f3}
\end{center}
\end{figure}

In Fig. \ref{f3}(a) we compare the signals when there is no puddle (red curve) in the system with the one when an e-h puddle is present (blue curve) with ($x_p, y_p, r_p$) = (0.48, 0.75, 0.05). The two signals differ in the region of densities between $n_1$ and $n_2$. Therefore, this is the region affected by the e-h puddle. The snake trajectory for density $n_1$ is given schematically in the inset of Fig. \ref{f3}(a). In order to find the size of the puddle underneath the gated region we need to calculate the distance that every part of the electron beam on the n-side (small orbit) covers when we increase the density below the top-gate voltage from $n_1$ to $n_2$. This is given by,
\begin{equation}\label{e4}
d_p = \frac{r_{TG}(n_2) - r_{TG}(n_1)}{2n_c} - 2r_{BG},
\end{equation}
where $d_p$ stands for the diameter of the puddle and $n_c$ is the number of orbits that the beam forms on the p-side below the puddle's position. In Fig. \ref{f3}(a) $n_c = 2$ because there are only two orbits below the center of the puddle. The second term in Eq. \eqref{e4} is a correction because at point $n_1$ the beam touches the puddle however at point $n_2$ the whole orbit on the n-side needs to be outside the puddle. Using this approach we get for the system of Fig. \ref{f3}(b) $d_p = 0.108$. Fig. \ref{f3}(b) shows the results for $B = 10$ and ($x_p, y_p, r_p$) = (0.49, 1.25, 0.035). In this case $n_c = 3$ because the beam executes three orbits below $y_p = 1.25$ and using Eq. \eqref{e4} this results in $d_p = 0.076$.

Alternatively, we can also estimate the puddle size by varying the magnetic field instead of the carrier density. For example, we decrease the cyclotron radius in the gated region by lowering the density in that region and by changing the magnetic field. The idea is to have a small cyclotron radius in the gated region and a large cyclotron radius in the non-gated region. The reason to do so is to obtain a better estimate of its size on the p-side of the junction. We proceed as in the previous case and search for the region where the two responses differ. However, in this case the cyclotron radii in both regions are changing with the magnetic field and therefore we need to change Eq. \eqref{e4} so it will also include a change of $r_{BG}$. In this case we need to sum all cyclotron orbits below $y_p$ for points $B_1$ and $B_2$ and subtract them to get the diameter of the puddle. This is shown in Fig. \ref{f3}(c) from which we derive $d_p = 0.115$. 

In Fig. \ref{f3}(d) we show another approach. Here, we increase the cyclotron radius in both regions to values much larger than the expected size of the puddle. The idea is to get rid of the correction from Eq. \eqref{e4}. Calculated diameter is $d_p = 0.09$. Notice that the previous methods always overestimate puddle's diameter which is due to the small radius on one side of the junction, however using the last method one arrives at an estimate which is smaller than the actual size of the puddle and thus in the end we realize a lower and upper bound for the size of the puddle. 

In the case when there is more than one puddle, disentangling the effects of each puddle can be very difficult. However, in certain cases it can be done. For example, in the case of two puddles we will be still able to determine the position of both puddles. Lets say we have two puddles at positions $y_1$ and $y_2$ and $y_2 > y_1$, where we used coordinate system as in Fig. \ref{f0}(a). Furthermore, lets assume that the direction of current is the same as in this figure. We showed that for determining the position effective length of the p-n interface $L^*_{PN} = W - y_p$ is important, i.e. the length from the puddle to the end of the p-n interface, since the new features  satisfy Eq. \eqref{entg} with $L_{PN} = L^*_{PN}$. In the case of two (or more) puddles this will be the length from the puddle that is situated closest to the opposite end of the p-n interface with respect to the current flow direction, in our case $y_2$ and we have $L^*_{PN} = W - y_2$. However, if we flip the direction of the current by e.g. changing the direction of the magnetic field, $L^*_{PN}$ will also change because now the effective length of the p-n interface will be $L^*_{PN} = y_1$. In this way, we are able to obtain $y_1$ and $y_2$.

In this paper we focused on the e-h puddles that are near the p-n interface. However, using this technique we are also able to determine the position of the e-h puddle formed further away from the p-n interface. The procedure is as follows. Lets assume that a puddle is formed at distance $d_p$ from the p-n interface. We can set parameters such that the cyclotron radius in both regions of the p-n interface are small. Then we start increasing the cyclotron radius by e.g. increasing the Fermi energy. We know that the conductance will be affected by the puddle only for $r_c \geq d_p$, i.e. when $r_c$ is large enough for electrons to reach the e-h puddle. Hence, by determining $d_p$ we obtain $x_p$ as $x_p = D - d_p$. On the other hand, procedure to obtain $y_p$ is the same as in the case of e-h puddle close to the p-n interface.
\section{Conclusion}

We investigated the influence of an e-h puddle placed close to the p-n interface on the transport properties of a graphene p-n junction in the presence of a magnetic field. The puddle alters the oscillatory behaviour of $dG/dV_{TG}$ versus magnetic field and new features such as crossings between different maxima and oscillations with different period appear. We showed that these features can be used to extract information about the position and the size of the puddle. Namely, from the period of the new oscillations one can obtain the position of the puddle, while the puddle's size can be estimated from a careful manipulation of the cyclotron radius in the two regions of the p-n junction.
\section*{Acknowledgement}

This work was supported by the Flemish Science Foundation (FWO-Vl) and
the European Science Foundation (ESF) under the EUROCORES Program
EuroGRAPHENE within the project CONGRAN. We acknowledge interesting correspondence with Thiti Taychatanapat.
\end{document}